# On virtual phonons, photons, and electrons

Günter Nimtz*
II. Physikalisches Institut, Universität zu Köln
Zülpicherstrasse 77, 50937 Köln
(Dated: October 24, 2018)

A macroscopic realization of the strange virtual particles is presented. The classical Helmholtz and the quantum mechanical Schrödinger equations are analogous differential equations. Their imaginary solutions are called evanescent modes in the case of elastic and electromagnetic fields. In the case of non-relativistic quantum mechanical fields they are called tunneling solutions. The imaginary solutions of this differential equation point to strange consequences: They are non local, they are not observable, and they described as virtual particles. During the last two decades QED calculations of the imaginary solutions have been experimentally confirmed for phonons, photons, and for electrons. The experimental proofs of the predictions of the non-relativistic quantum mechanics and of the Wigner phase time approach for the elastic, the electromagnetic and the Schrödinger fields will be presented in this article. The results are zero tunneling time and an interaction time (i.e. a phase shift) at the barrier interfaces. The measured barrier interaction time (i.e. the barrier transmission time) scales approximately inversely with the particle energy.

## I. INTRODUCTION

The effective interaction between electrons due to the exchange of virtual phonons gives rise to superconductivity. In optics evanescent modes, i.e. virtual photons became known from total reflection. If the transmitted beam is refracted $\geq 90^0$ in the optical lesser dense material, total reflection takes place as follows from Snell's law

$$n_1 sin\alpha = n_2 sin\beta, \qquad (1)$$

Such a behavior of the transmitted wave is illustrated in Fig. 1, where $n_1 > n_2$ is assumed for the refractive indices. We get for the wave number the relation[1]

$$k_x = \sqrt{\frac{\omega^2}{c^2}(1 - (n_1/n_2)^2 sin^2\alpha)}, \qquad (2)$$

where $k_x$ is the transmitted component of the wave vector in the direction normal to the surface (see Fig. 1), and $\omega$ and c are the angular frequency and the velocity of light, respectively. For $(n_1/n_2)^2 sin^2\alpha > 1$ the wave number $k_x$ becomes purely imaginary and the field drops off exponentially with increasing x, the field has become evanescent.

By introducing a second optical dense material from the right side of the sketch the total reflection becomes frustrated: The wave number of the evanescent mode becomes real again in the dense medium and the total reflection is reduced by the amount of the transmitted beam. This frustrated total internal reflection (FTIR) is illustrated with the typical double prisms in Fig. 2. Part of the incident beam is transmitted through the gap of length d. The transmission increases exponentially with decreasing gap. This symmetrical design of the beam path in Fig. 2 allows a direct measurement of the time a signal spends crossing the gap d.

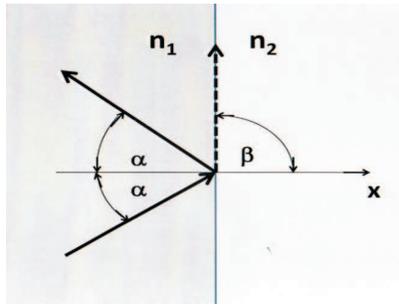

FIG. 1: Sketch of total reflection of a beam at a boundary of two media with the refractive indices $n_1 > n_2$.

Theoretical investigations resulted in zero time tunneling, i.e. the time spent inside an opaque barrier is zero[2,3,4,5,6,7]. Opaque means that the exponent of the wave function or of the evanescent mode is given by $\kappa x \geq 1$, where $\kappa$ and x are the imaginary wave number inside the barrier and the barrier length, respectively. A recent study suggests zero time tunneling as means to test the notional time translation theory[8].

There are still several studies and discussions published calculating a finite time spent inside a barrier and/or denying the possibility of superluminal signal velocity, for instance Refs.[9,10,11,12,13]. The discussion goes on in spite of the theoretical investigation by Collins et al.: *The quantum mechanical tunnelling time problem - revisited*. Their rigorous investigation concluded that the early quantum mechanical calculation by Hartman and the phase time approach present the correct approach to describe tunneling time[7]. Experimental studies have confirmed this statement [14,15,16]. A short barrier transmission time is simultaneously spent at the boundaries of a barrier, whereas zero time is actually spent inside the barrier[16].

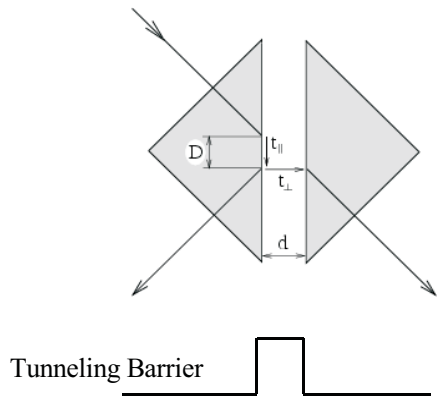

FIG. 2: Sketch of the time components in the investigated symmetrical beam design of double prisms under the condition of FTIR. Reflected and transmitted signals were detected at the same time in spite of the fact that the transmitted beam traveled an additional distance d through the gap. An analog tunneling square potential barrier is sketched at the bottom[14,17].

The experimental design in which the two time contributions can easily be distinguished is illustrated in Fig. 2 [14]. The time component $t_\parallel$ arises at the barrier front and $t_\perp$ inside the barrier as seen in Fig. 2. It holds $t_\perp = 0$, since the measured reflection time and transmission time of the signals are equal. Thus we get for the transmission (i.e. tunneling) time of photons

$$\tau = t_\parallel + t_\perp = t_\parallel \qquad (3)$$

Zero tunneling time for electrons was calculated in Refs.[2,5] for example and recently expected from electron tunneling experiments presented in Ref.[19].

The transmission time $\tau$ measured in various experiments was observed to be of the order of magnitude of the reciprocal frequency T of the wave packets [18]. Obviously the tunneling time scales with the inverse energy of the wave packet.

$$\tau \approx \frac{1}{\nu} = T, \qquad (4)$$

where $\tau$ is the measured barrier transmission time and $\nu$ is the carrier frequency of the electromagnetic wave packet. Incidentally, D the Goos-Hänchen shift of FTIR in Fig.[14] equals approximately one wave length[18].

In the case of a wave packet with a rest mass the transmission time is approximately

$$\tau \approx \frac{h}{W_{kin}}, \qquad (5)$$

where h is the Planck constant and $W_{kin}$ is the particle energy.

The symmetrical double prisms set-up represents the analog to quantum mechanical tunneling of a square potential barrier as was pointed out by Sommerfeld[17].

The observed tunneling time behavior for certain special barriers provided a theoretical basis for Esposito[20]. He deduced the modified relation

$$\tau_A = \frac{1}{\nu} \cdot A, \tag{6}$$

where the factor A depends on the specific barrier and wave packet in question[20,21]. Recent experimental data on phonon and electron tunneling have pointed to a similar barrier traversal time relation ship[21,22].

## II. WAVE PROPAGATION: ONE DIFFERENTIAL EQUATION

For sound and electromagnetic waves the propagation can be described by the Helmholtz and the Maxwell equations and for massive particles by the Klein–Gorden, the Dirac or, in the non–relativistic regime, by the Schrödinger equation.

The Maxwell equations in media are characterized by a refractive index $n = \sqrt{\mu \epsilon}$, where $\mu$ and $\epsilon$ are the relative permeability and the relative permittivity, and lead to the wave equation

$$-\nabla^2 \phi(\vec{x}, t) + \frac{n^2}{c^2} \frac{\partial^2}{\partial t^2} \phi(\vec{x}, t) = 0, \tag{7}$$

$\phi$ being any component of the electrical and the magnetic fields or an elastic field. In vacuum characterized by $n = 1$ waves propagate with the velocity $c = (\mu_0 \epsilon_0)^{-1/2}$, where $\mu_0$ and $\epsilon_0$ are the permeability and the permittivity, respectively.

If we describe phenomena periodic in time with frequency $\nu = \omega/(2\pi)$,

$$\phi(\vec{x}, t) = \phi_x(\vec{x}) e^{i\omega t}, \tag{8}$$

then the wave equation reduces to the Helmholtz equation

$$\nabla^2 \phi_x(\vec{x}) + \frac{n^2 \omega^2}{c^2} \phi_x(\vec{x}) = 0. \tag{9}$$

As usual, this equation will be solved by a plane wave Ansatz

$$\phi_a(\vec{x}) = \phi_0 e^{-i\vec{k} \cdot \vec{x}}, \tag{10}$$

and we get a relation between the wave number and the refractive index

$$k^2 = \frac{n^2 \omega^2}{c^2} = k_0^2 n^2 = k_0^2 \epsilon \mu, \tag{11}$$

where $k_0$ is the wave number in free space. If $k$ and then $n$ are imaginary numbers, the solution is called an evanescent mode.

In the case of an elastic field the wave number is

$$k^2 = \frac{\omega^2}{v_s^2}, \tag{12}$$

where $v_s$ is the velocity of sound. The latter depends on the elasticity modulus and on the mass density.

Similar features can be found for the stationary Schrödinger equation

$$W_{kin} \psi(\vec{x}) = -\frac{\hbar^2}{2m} \nabla^2 \psi(\vec{x}) + U(\vec{x}) \psi(\vec{x}), \tag{13}$$

where $W_{kin}$ is the energy of the stationary state, $m$ is the mass of the particle and $U(\vec{x})$ is a position–dependent potential, the barrier potential, for example. This relation is mathematically equivalent to the Helmholtz equation

$$\nabla^2 \psi(\vec{x}) + \frac{2m}{\hbar^2} (W_{kin} - U(\vec{x})) \psi(\vec{x}) = 0. \tag{14}$$

Again, a plane wave Ansatz

$$\psi(\vec{x}) = \psi_0 e^{-i\vec{k}\cdot\vec{x}} \qquad (15)$$

yields for the wave number $k$

$$k^2 = \frac{2m}{\hbar^2}(W_{kin} - U) = k_0^2 - \frac{2mU}{\hbar^2}, \qquad (16)$$

where $k_0^2 = 2mE/\hbar^2$ is the wave vector at infinity, where $U$ is assumed to vanish. Particles in regions for which $W_{kin} < U$, that is, inside the potential barrier, are quantum analogues of evanescent modes. The refractive index plays the role of the potential for the electromagnetic evanescent modes in wave mechanical tunneling.

The relevant quantities for sound modes are the elasticity modulus and the mass density.

## III. EXPERIMENTAL DATA

Analogous tunneling was first studied with digital microwave pulses in undersized wave guides by Enders and Nimtz in 1992[28]. The result was a superluminal tunneling velocity, The superluminal velocity has been reproduced in experiments at microwave, infrared, and optical frequencies in photonic band gap material, i.e. optical mirrors, and in double prisms FTIR, see, for instance, the reviews Refs.[16,29]. Ten years ago, Peter Mittelstaedt made a statement on these macroscopic experiments: *That is quantum mechanics in the living room*[30]. The studied double prisms were cut from a large perspex cube of the size 0.4 x 0.4 x 0.4 $m^3$ and the microwave length was 3 cm, see Ref.[14].

The barrier transmission time was obtained in the optical analogous experiments either by directly measuring the barrier transmission time (time domain) or by calculating the corresponding time by the phase time approach (frequency domain)

$$\tau = \hbar d\varphi/dW_{kin} \qquad (17)$$
$$\tau = -d\varphi/d\omega, \qquad (18)$$

where $W_{kin}$ is the energy, $\varphi$ is the phase shift of the wave at the barrier, and $\omega$ is the angular frequency; $\varphi$ is given by the real part of the wave number k times the distance x. In the case of evanescent modes and of tunneling solutions the wave number k is purely imaginary. Thus propagation of evanescent or tunneling modes appears to take place in zero time[5].

As mentioned above zero time tunneling of wave packets as well as the barrier interaction time become most obvious in FTIR with double prisms. The symmetrical design of the beam paths and the corresponding traveling time components are sketched in Fig.2. The experimental result revealed for the traversal time $t_\perp = 0$, because reflected and transmitted signals arrived at the detectors at the same time. The Goos-Hänchen shift of about one wavelength is related to the interaction time at the barrier boundaries. The latter is also valid for barriers like forbidden band gaps of periodic dielectric hetero structures.

The universal tunneling time relation appears to be valid also for sound waves. The observation is reasonable since elastic and electromagnetic fields are described by the same differential equations. Yang et al., for instance, investigated the tunneling time of ultrasonic pulses in the forbidden frequency gap of a phononic crystal. Ultrasonic pulses had a carrier frequency of 1 MHz. They measured a transmission time between 0.6 and 1.0 $\mu s$[25]. Similar acoustical experiments at a frequency of 1 kHz by Robertson et al. obtained a transmission time of about 1 ms [26].

An electron tunneling time of 7 fs was measured by a field-emission microscopic experiment[36]. The barrier height was 1.7 eV, assuming an energy of 0.7 eV we get 6 fs by Eq. 5 An extremely short atom ionization electron tunneling time of some attoseconds or actually zero time was reported recently[19]. Equation 5 would give a transmission time of 76 as assuming $W_{kin} = 54.39$ eV, which is the ground state energy of the electron in helium. However, this special tunneling process can only be qualitatively compared with one dimensional free particle tunneling.

## IV. DISCUSSION

In 1962 Hartman's quantum mechanical study on tunneling of a wave packet for opaque barriers revealed a superluminal velocity and a tunneling time independent of barrier length[2,31]. The calculations were confirmed by the above mentioned photonic tunneling experiments, see for instance Refs.[31,32,35,37]. Haibel and Nimtz, figured out that the tunneling time of opaque barriers scales in first order approximation with the reciprocal frequency of the wave packet,[18]. Experimental data for phonons, photons, and electrons are presented in table 1. In the case of simple one-dimensional tunneling barriers the empirical relationships (Eqs. 4,5) represent a good approximation.

| Tunneling barriers | Reference | $\tau$ | $T = 1/\nu$ |
|---|---|---|---|
| *frustrated total reflection* | Ref.[18] | 117 ps | 120 ps |
| *at double prisms* | Ref.[23] | 30 fs | 11.3 fs |
| | Ref.[24] | 87 ps | 100 ps |
| *photonic lattice* | Ref.[32]. | 2.13 fs | 2.34 fs |
| | Ref.[37] | 2.7 fs | 2.7 fs |
| *undersized waveguide* | Ref.[28] | 130 ps | 115 ps |
| *electron field-emission tunneling* | Ref.[36] | 7 fs | 6 fs |
| *electron ionization tunneling* | Ref.[19] | $\leq 6$ as | ? as |
| *acoustic (phonon) tunneling* | Ref.[25] | 0.8 $\mu$s | 1 $\mu$s |
| *acoustic (phonon) tunneling* | Ref.[26] | 0.9 ms | 1 ms |

TABLE I: Experimental and $1/\nu$ transmission time (*tunneling time*), of photons, phonons, and electrons. Notice, there are about 15 orders of magnitude in different barrier traversal times.

In the abstract three strange properties of evanescent and tunneling modes are mentioned: they are non local, tunneling particles are not observable, and they are virtual particles.

### A. Evanescent modes are not local

Experiments have shown that evanescent modes are present at the same time all over the barrier, i.e. at the front and at the back interface of the barriers.

### B. Evanescent modes are not observable

Remarkably, evanescent modes like tunneling particles are not observable inside a barrier[33,34].

The electric energy density $u$ of the evanescent electric field $E$ with its imaginary refractive index is negative:

$$u = \frac{1}{2}\epsilon E^2 < 0, \tag{19}$$

$$\epsilon_r = n^2 < 0, \tag{20}$$

where $\epsilon = \epsilon_0 \cdot \epsilon_r$

In the case of particle tunneling we get a negative total energy $W$

$$W = W_{kin} - U < 0, \tag{21}$$

where $W_{kin}$ and $U$ are the kinetic energy and the potential barrier height, respectively. Equations 19, 21 are the quantum electrodynamic basis for the existence of evanescent waves.

An evanescent field does not interact with real fields due to the imaginary wave number resulting in an impedance mismatch. Fields can only transmit energy if the reflection at an antenna $R < 1$ holds. If $n_1$ represents the purely imaginary refractive index of an evanescent region and $n_2$ represents the complex refractive index of the receiver medium then the square of the absolute value gives

$$R = |r|^2 = \frac{|n_2 - n_1|^2}{|n_2 + n_1|^2}, \tag{22}$$

which equals 1 and total reflection takes place. In order to account for evanescent modes they have to be transmitted into medium, where they have a real wave number.

In order to observe a particle in the exponential tail of tunneling probability, it must be localized within a distance of order $\Delta x \approx 1/\kappa$, where $\kappa$ is the imaginary wave number in question. Hence, its momentum $\Delta p$ must be uncertain in the cases of Schödinger particles and FTIR by

$$\Delta p \;>\; \hbar/\Delta x \approx \hbar\kappa = \sqrt{2m(U - W_{kin})} \qquad (23)$$

The particle of positive energy $W_{kin}$ can thus be located in the nonclassical region only if it is given an energy $U - W_{kin}$, sufficient to raise it into the classically allowed region[33,34]. In this case $\Delta$p tends to 0 and p becomes measurable.

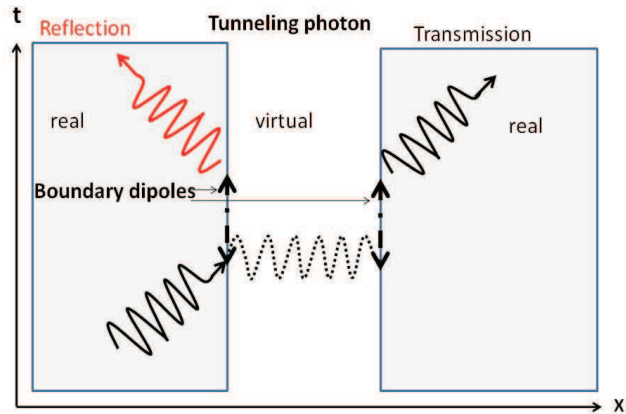

FIG. 3: FTIR laboratory experiments result in this t-x diagram of the interaction mechanisms of photon tunneling[14,16]. The double-headed arrows represent the dipoles at the two interfaces prism-air-prism. The interaction time is simultaneously raised at the two interfaces by the virtual photon inside the barrier. This behavior is in agreement with the Hartman effect, which states that the traversal time is independent of the barrier length[2,35]. Accordingly reflected and transmitted photons display the same barrier interaction time. .

### C. Evanescent modes are virtual photons

In the early 1970's, QED calculations showed that evanescent modes are adequately described by virtual photons[27]. The quantization of evanescent modes by Carniglia and Mandel showed that the locality condition is not fulfilled[3,27]. They concluded that the commutator of the field operator does not vanish for space-like separated points. Thus an evanescent mode violates the microscopic causality. It was mentioned by Ali[4] that virtual photons are those modes, which do not satisfy the Einstein relation

$$W^2 \;=\; (\hbar k)^2 c^2. \qquad (24)$$

A t-x graph of the photon interaction mechanisms with the barrier interfaces in the tunneling process of a photon is shown in Fig. 3. The simultaneous interaction is performed by the boundary dipoles in the dielectric media.

### D. Virtual phonons and electrons

Tunneling phonons i.e. bosons are behaving like photons. Tunneling electrons are according to quantum mechanics also virtual particles Refs.[33,34]. As mentioned above, recent experimental electron tunneling data point like the photon experiments to zero time tunneling.

### E. Summary

Evanescent and tunneling modes propagate in zero time. They are presented by virtual particles. This contradicts the often given interpretation of tunneling by the uncertainty relation, where the particle may acquire energy to overcome the barrier. However, this would cause a comparatively long vacuum barrier traversal time, much longer than the observed traversal time, which arise due to the two interfaces interaction. As a matter of fact the tunneling process is independent of the special physical meaning of the wave considered, it is the same for elastic, electromagnetic and Schrödinger waves.

---